\documentclass[aps,floatfix]{revtex4}
\usepackage{graphics,epsfig}
\usepackage{graphicx}

\begin{document}

\title{Five-dimensional bulk with a time-dependent warp factor and its consequences on brane cosmology}

\author{Sarbari Guha$^1$ and Subenoy Chakraborty$^2$}
\address{$^1$ Department of Physics, St. Xavier's College (Autonomous), Kolkata 700 016, India}
\address{$^2$ Department of Mathematics, Jadavpur University, Kolkata 700032, India}

\begin{abstract}
In this paper, we have considered a 5-dimensional warped product space-time with a time-dependent warp factor. The time-dependent warp factor plays an important role in localizing matter to the 4-dimensional hypersurface constituting the observed universe. We then proceed to determine the nature of modifications produced in the bulk geometry as well as the consequences on the corresponding braneworld. The five-dimensional field equations are constructed. For the bulk metric chosen, the Weyl tensor is found to vanish under a certain condition, thereby satisfying the conditions of a constant curvature bulk. Consequently we have constructed the solutions to the field equations for such a bulk and have studied its physical properties. The braneworld is described by a flat FRW-type metric in the ordinary spatial dimension. It is found that the the effective cosmological constant for the induced matter is a variable quantity monitored by the time-dependent warp factor and is associated with the bending of the braneworlds. The universe is initially decelerated, but subsequently makes a transition to an accelerated phase at later times, when its dynamics is dominated by the dynamical dark energy component. For a particular choice of the warp factor, it has been found that the universe is represented by a model of dark energy with the age of the universe as the measure of length, driving the universe to a state of accelerated expansion at later times.

\end{abstract}

\maketitle

\section{Introduction}
\label{intro}
Theories of extra dimensions have aroused the interest of physicists ever since the works of Kaluza and Klein \cite{kk}. Although these extra dimensions might have played an important role in the evolution of the early universe, but till now they have remained undetected in experiments at energy levels within the TeV scale. The immediate question that one has to address then is how to explain the 4-dimensionality of the observed universe. There are theories seeking to explain the reason for such non-observability \cite{rbvspv}. In spite of this, extra-dimensional theories continue to attract wide attention since these have been successfully used to solve the hierarchy problem of particle physics \cite{aadd}\cite{add} and to explain the idea that the post-inflationary epoch of our universe was preceded by the collision of D3-branes \cite{khoury}. This idea led to the so-called 'braneworld scenario', where the ordinary standard model matter and non-gravitational fields are confined by some trapping mechanism to the 4-dimensional universe constituting the D3-branes (4-dimensional timelike hypersurfaces), embedded in a $(4 + n)$-dimensional 'bulk' (n being the number of extra dimensions). At low energies, gravity is confined to the brane along with particles, but at high energies gravity "leaks" into the higher-dimensional bulk, so that only a part of it is felt in the observed 4-dimensional universe. The strong constraints on the size of extra dimensions as in the Kaluza-Klein models, is therefore relaxed \cite{add}, lowering the fundamental Planck scale down to the TeV range. The main role of the extra dimensions in such theories is to provide extra degrees of freedom on the brane.

Among the several higher-dimensional models developed over the years, the one which has turned out to be popular for a number of reasons, is the warped braneworld model of Randall and Sundrum \cite{rs1}\cite{rs2}, with a single extra dimension. In their model, matter fields were considered to be localized on a 4-dimensional hypersurface in a constant curvature five-dimensional bulk furnished with mirror symmetry, with an exponential warp factor and a non-factorizable metric, even when the fifth dimension was infinite. The field equations on the corresponding 4-dimensional universe, was modified by the effect of the extra dimension. In their case, the warp factor was a function of the extra coordinate and the metric for the extra-dimensional coordinate was constant. However, both of these parameters can be functions of time and the extra coordinate. In the process, the solutions get very much complicated, and in many cases cannot be determined without the imposition of suitable constraints.

In this paper, we have considered RS-type braneworlds with the bulk in the form of a five-dimensional warped product space-time, having an exponential warp factor which depends both on time as well as on the extra coordinates and a non-compact fifth dimension. We know that the exponential warp factor reflects the confining role of the bulk cosmological constant \cite{lrr} to localize gravity at the brane through the curvature of the bulk. It is, therefore, possible that such a localization may include some time-dependence and so we make such a choice. It is then necessary to determine the modifications produced in the bulk as well as the consequences on the corresponding braneworld. The 5-dimensional bulk is assumed to have constant curvature. Although such a bulk has limited degrees of freedom, yet it is sufficient for us if the braneworld is of FRW-type, in which case no additional conditions are necessary to compensate for the reduced degrees of freedom. The advantage of assuming the braneworld to be of FRW-type is that it can be embedded in any constant curvature bulk including the de Sitter $dS_{5}$, anti de Sitter $AdS_{5}$ and the flat Minskowski manifold \cite{maia1}\cite{maia2}. Therefore, the braneworld considered by us is assumed to be defined by a flat FRW-type metric in the ordinary spatial dimension.

The plan of the paper is as follows: In section II, the basic theoretical framework necessary for us, has been discussed. This has been followed by the construction of the field equations in the next section. The energy scales have been specified to impose constraints on the bulk geometry and subsequently the field equations are simplified. In section IV, the solutions to the field equations are constructed for a constant curvature bulk and the effect of the warp factor on the bulk geometry as well as on the energy conditions in the bulk has been analysed. In section V, the evolution of the observed universe has been studied. The effect of the time-dependent warp factor in modifying the usual dynamics of the gravitational field compared to that predicted from Einstein's theory, has been discussed. The vacuum energy of the corresponding braneworld is found to be a time-varying entity associated with the bending of the braneworlds. In section VI, the cosmological implications for such a braneworld has been analysed. The summary of the entire exercise has been presented in Section VII.

\section{Theoretical considerations}
\label{sec:1}

Let us consider a 5-dimensional theory with the action decomposed as \cite{Nihei}
\begin{equation}\label{01}
S= -\frac{1}{2\kappa^{2}_{(5)}}\int d^{5}x \sqrt{\bar{g}}[\bar{R}+2\Lambda_{(5)}] + \int d^{4}x \sqrt{-\bar{g}}L_{m}
\end{equation}
where, $\bar{g}_{AB}$ is a 5-dimensional metric of signature (+ - - - -). Here $\Lambda_{(5)}$ is the bulk cosmological constant and $\bar{R}$ is the 5-dimensional scalar curvature for the metric. The first term in (\ref{01}) corresponds to the Einstein-Hilbert action in 5-dimensions. The second term represents the four-dimensional boundary action localized at $y = 0$, the Lagrangian density $L_{m}$ representing all contributions to the action, which are not strictly gravitational. The constant $\kappa_{(5)}$ is related to the 5-dimensional Newton's constant $G_{(5)}$ and the 5-dimensional reduced Planck mass $M_{(5)}$ by the relation
\begin{equation}\label{02}
\kappa^{2}_{(5)}= 8\pi G_{(5)}= M^{-3}_{(5)}.
\end{equation}
The bulk manifold is assumed to be a space of constant curvature $\bar{K}$, so that the magnitude of the bulk Riemann tensor is determined by the curvature radius '$l$' of the bulk manifold as follows:
\begin{equation}\label{03}
\bar{R}_{ABCD}= \bar{K}[\bar{g}_{AC}\bar{g}_{BD} - \bar{g}_{AD}\bar{g}_{BC}]= \frac{\epsilon}{l^{2}}[\bar{g}_{AC}\bar{g}_{BD} - \bar{g}_{AD}\bar{g}_{BC}].
\end{equation}

For $AdS_{5}$ geometry, $\epsilon=-1$, whereas for $dS_{5}$, $\epsilon=1$. The bulk cosmological constant is related to the curvature $\bar{K}$ by the relation $\Lambda_{(5)}= 6\bar{K}$. The 5-dimensional field equations for such a bulk are read as \cite{lrr}\cite{bdl}
\begin{equation}\label{04}
\bar{G}_{AB} = - \Lambda_{(5)}\bar{g}_{AB} + \kappa^{2}_{(5)}\bar{T}_{AB}
\end{equation}
where $\bar{G}_{AB}$ is the 5-dimensional Einstein tensor and $\bar{T}_{AB}$ represents the 5-dimensional energy-momentum tensor.

We shall consider five-dimensional warped metrics with time-dependent warp factor in the form
\begin{equation}\label{05}
dS^{2}=e^{2f(t,y)}\left(dt^2-bt(dr^2+r^2d\theta^2+r^2sin^{2}\theta d\phi^2)\right)-h(t,y)dy^2.
\end{equation}
where $y$ is the coordinate of the fifth dimension, $t$ denotes the conformal time, $b$ is a constant and $h(t,y)$ is the scale factor representing the expansion of the orbifold, which in this case turns out to be a field parameter, called "radion". We assume that the fifth dimension is non-compact and curved (i.e. warped) \cite{lrr},\cite{rs2}. The smooth function $f$ is called the "warping function" and $e^{2f(t,y)}$ is the warp factor. As in the RS 1-brane models, in the present case also, gravity is localized on the brane through the curvature of the bulk. The single brane, has positive tension, which offsets the effect of the negative bulk cosmological constant. The bulk cosmological constant acts to "squeeze" the gravitational field closer to the brane, with the exponential warp factor reflecting the confining role of the bulk cosmological constant. We assume the warp factor to be a function of both time, as well as of the extra coordinate \cite{GC}. Mathematically, the time dependence of the warp factor does not affect the smooth nature of the function $f$. Physically, it takes into account the possibility that such a process of localization (i.e. the confining role of the bulk cosmological constant and the curvature of the bulk) may have some dependence on time, in addition to their dependence on the extra dimensional coordinate. The location of the hypersurface will then have an additional time-dependence together with a $y$-dependence. The $y$-dependence may give rise to bending effects on the hypersurface. Having considered a time-dependent process of localization of gravity, it is now necessary to determine the nature of modifications produced in the bulk geometry as well as the consequences on the corresponding braneworld.

For reasons already explained, we consider the hypersurface to be defined by a flat FRW-type metric in the ordinary spatial dimension. For the chosen metric, the ordinary 4-dimensional universe resembles a spatially flat Friedmann model with the scale factor evolving as half-power of the cosmic time and hence is in the radiative phase of evolution. However, the \emph{induced metric} has a scale factor $A$ which evolves as a composite function of conformal time, given by $A^{2}(\eta(t))=bte^{2f(t,y=0)}$, the observed universe being represented by the hypersurface $y=0$. Here, $\eta(t)\equiv\int dt e^{f(t,y=0)}$ is the proper time of a comoving observer, when the position of the hypersurface is fixed at $y=0$. The geometry of the observed universe at the location, $y=0$, will be determined by the induced metric

\begin{equation}\label{05a}
ds^{2}=g_{\alpha\beta}(x,y=0)dx^{\alpha}dx^{\beta}=e^{2f_{0}}q_{\alpha \beta}(x)dx^{\alpha}dx^{\beta},
\end{equation}
where $f_{0}$ is the value of $f$ at $y=0$ and $q_{\alpha\beta}=q_{\alpha\beta}(x)$ is the warp metric on the 4-dimensional hypersurface.

In the case where the brane is allowed to "bend", the location of the hypersurface will have some $y$-dependence. Geometrically, the extrinsic curvature of the hypersurface gives us a measure of the deviation of the hypersurface from the tangent plane and therefore such bending may produce an observable result in the form of a smooth scalar function represented by the warp factor \cite{maia4}. The extrinsic curvature is defined by
\begin{equation}\label{05b}
K_{\alpha\beta}=-\frac{1}{2}\pounds_{n}g_{\alpha\beta}(x^{\mu},y),
\end{equation}
where,  $x^{\mu}$ are the coordinates on the hypersurface and
\begin{equation}\label{05c}
K_{\alpha\beta;\gamma} - K_{\alpha\gamma;\beta}=0.
\end{equation}
The extrinsic curvature of this hypersurface \cite{dahia3},\cite{dahia4} is then given by
$K_{\alpha\beta}=-\frac{1}{2}\left(\frac{\partial g_{\alpha\beta}(x,y=0)}{\partial y}\right)$. In Section IV, we shall explore a possible relation between the extrinsic curvature of the hypersurface and vacuum energy-momentum tensor.

From the physical point of view, in brane models, the stress tensor for a source on the brane has a vanishing $yy$-component, $y$ being the extra-dimensional coordinate. It appears that one degree of freedom, in the form of a scalar field, decouples from the sources on the brane, thereby reproducing the ordinary massless 4-dimensional graviton propagator. It is known that the introduction of a source term on the brane cause the brane to bend in a frame in which the gravitational fluctuations are small \cite{garriga} \cite{giddings}. This bending exactly compensates for the effect of the additional 4D scalar field contained in the 5D graviton propagator, and so we recover the ordinary 4D massless graviton propagator in the RS model. In the present paper, we intend to study the effect of the time-dependent warp factor on the extrinsic curvature of the brane and interpret the consequences from the physical point of view.

To illustrate our study we choose to work with a specific example for the warp factor. Let the scalar function $f(t,y)$ be given by the simple form
\begin{equation}\label{06}
f(t,y)=\frac{1}{2}(\ln\tau(t)+\ln\Gamma(y)).
\end{equation}

With this choice, the five-dimensional metric assumes the form
\begin{equation}\label{07}
dS^{2}=\tau(t)\Gamma(y)\left(dt^2-bt(dr^2+r^2d\theta^2+r^2sin^{2}\theta d\phi^2)\right)-h(t,y)dy^2.
\end{equation}
The scale factor for the induced metric is now $A^{2}=bt\tau(t)e^{\ln \Gamma(y=0)}=Cbt\tau(t)$ where $C=e^{\ln \Gamma(y=0)}=\Gamma(y=0)$. The nature of the constraints on the functions $\tau$, $\Gamma$ and $h$ will determine whether the bulk will be a de Sitter or an Anti de Sitter one. We shall take up specific cases as we progress.

The components of the bulk stress-energy tensor are

\begin{center}
$\bar{T}^{t}_{t}=\bar{\rho},\qquad\qquad \bar{T}^{i}_{j}=-\bar{P},\qquad\qquad \bar{T}^{y}_{y}=-\bar{P}_{y}$.
\end{center}

In theories of extra dimensions, it has been shown that the radius $R_{y}$ of the extra dimensions where the gauge interactions propagate, can be large enough \cite{ant1} for the corresponding excitations to be at the reach of future accelerators \cite{ant2}, while the string scale $M_{p}$ can be lowered to the TeV range \cite{Lykken}. The effective strength of gravity is related to the volume of the extra dimensions \cite{aadd} \cite{add}. In that case, the effective four-dimensional Newton's constant $G_{N}$ is inversely proportional to the total volume of the extra dimensions, which in our case, is \cite{Chung}
\begin{equation}\label{07a}
G_{N}\sim \frac{\kappa^{2}_{(5)}}{16 \pi \Theta}
\end{equation}
where $\Theta$ is the physical length of the fifth dimension, given by
\begin{equation}\label{07b}
\Theta \equiv \int \sqrt{-g_{yy}} dy = \int h(t,y)^{1/2}dy.
\end{equation}
Any variation in the extra-dimensional volume will show up as a variation of $G_{N}$ \cite{clinevinet}
\begin{equation}\label{07c}
\frac{\dot{G}_{N}}{G_{N}}=-\frac{\dot{h}}{2h}.
\end{equation}
Consequently, for a stabilized bulk, we have
\begin{equation}\label{07d}
\dot{h}=0,
\end{equation}
so that the scale of gravity will not show any fluctuation. For such a case, we can determine the explicit dependence of the metric on the $y$ coordinate.

\section{Five-dimensional Field Equations}

The non-vanishing components of the five-dimensional Einstein tensor for the warped product spacetime given by (\ref{07}) are obtained as

\begin{equation}\label{08}
\bar{G}_{tt}=\frac{3\bar{g}_{tt}}{4\tau\Gamma}\left( \frac{1}{t} \left( \frac{2\dot{\tau}}{\tau} + \frac{\dot{h}}{p} + \frac{1}{t} \right) + \frac{\dot{\tau}}{\tau}\left( \frac{\dot{\tau}}{\tau} + \frac{\dot{h}}{h} \right) \right) + \frac{3\bar{g}_{tt}}{4h}\left( \frac{\Gamma^{\prime}h^{\prime}}{\Gamma h} - \frac{2\Gamma^{\prime \prime}}{\Gamma} \right)
\end{equation}

\begin{equation}\label{09}
\bar{G}_{ty}=\frac{3 \dot{h} \Gamma^{\prime} }{4 h \Gamma}
\end{equation}

\begin{equation}\label{10}
\bar{G}_{yy}=\frac{3\bar{g}_{yy}}{4\tau\Gamma}\left( \frac{2\ddot{\tau}}{\tau} + \frac{3\dot{\tau}}{t\tau} - \left( \frac{\dot{\tau}}{\tau} \right)^{2} \right) - \frac{6\bar{g}_{yy}}{4h}\left( \frac{\Gamma^{\prime}}{\Gamma} \right)^{2}
\end{equation}

\begin{equation}\label{11}
\bar{G}_{ij}=\frac{\bar{g}_{ij}}{4\tau\Gamma} \left( \frac{4\ddot{\tau}}{\tau} + \frac{\dot{\tau}}{\tau} \left( \frac{1}{t} + \frac{\dot{h}}{h} - \frac{3\dot{\tau}}{\tau} \right) + \frac{\dot{h}}{h} \left( \frac{2}{t} - \frac{\dot{h}}{h} \right) + \frac{2\ddot{h}}{h} - \frac{1}{t^{2}} \right) + \frac{3\bar{g}_{ij}}{4h}\left( \frac{\Gamma^{\prime}h^{\prime}}{\Gamma h} - \frac{2\Gamma^{\prime \prime}}{\Gamma} \right)
\end{equation}
Above, an overdot represents derivative with respect to time $t$ and a prime stands for a derivative with respect to the fifth coordinate $y$.

According to Arkani-Hamed \cite{aadd}\cite{add}, the standard model fields are localized within a distance $m^{-1}_{EW}$ in the extra dimension, where $m_{EW}$ is the electroweak scale. They acquire momentum in the extra dimension and escape from our 4-dimensional world, carrying energy, only in sufficiently hard collisions of energy $E_{esc}\gtrsim m_{EW}$. At lower energies, particles remain confined to the 4-dimensional universe. To prevent matter or energy escaping from the brane along the fifth dimension at low energies, we require $\bar{T}^{t}_{y}=0$, which implies that $\bar{G}^{t}_{y}=0$. Therefore a non-zero $\bar{G}^{t}_{y}$ term is necessary for us to be able to probe the extra dimensions. For the metric under our consideration, we find that $\bar{G}^{t}_{y}\neq0$. The geometry is therefore suitable for the standard model fields to be able to access the extra dimension. However, the energy scale necessary to ensure such an access is few hundred GeV. The energy is high enough to ensure the production of the first KK electron-positron pair. The bulk and brane-localized radiative corrections then give rise to the subsequent splitting of the heavy modes into the zero modes of the SM particles and the first KK photon. In the theory proposed by Randall and Sundrum \cite{rs1}\cite{rs2}, with only one additional dimension, the hierarchy between the fundamental five-dimensional Planck scale and the compactification scale was only of order 10. Consequently, the excitation scale was of the order of a TeV, there was no light KK mode and the other KK modes were expected to be detected at high energy colliders.

In the higher-dimensional theories with large extra dimensions, it has been proved that the higher-dimensional gauge invariance protects the Higgs mass from quadratic divergences at the quantum level and one-loop radiative corrections to the Higgs mass are finite (ultraviolet insensitive) and of the order of the inverse of the radius $R_{y}$ of the extra dimensions \cite{ant3}. All of them are based upon introducing a symmetry at an intermediate scale $M_{0}$ between the electroweak scale $M_{weak}$ and the Standard Model cutoff (string scale $M_{p}$ ) such that quadratic divergences are canceled at scales $\mu > M_{0}$. In this way quadratic divergences survive only for scales smaller than $M_{0}$ and radiative corrections to the Higgs mass are $\sim M_{0}$. For scales $\mu<1/R_{y}$, the theory is four-dimensional and it is not protected from quadratic divergences, while for $\mu > 1/R_{y}$ the theory is higher-dimensional and the gauge invariance protects the Higgs mass from quadratic divergences. In both cases, for the mechanism to be effective, the scale $M_{0}$ has to be stabilized and should be not much higher than the electroweak scale. Determining the radius $R_{y}$ implies that we are considering the gravitational sector of the theory involving the radion field. The corresponding mathematics is extremely complicated and it is often assumed that the radius has been fixed and stabilized by some mechanism \cite{ponton}.

To describe gravity at lower energy scales, we need $\bar{G}^{t}_{y}=0$.  For the metric under our consideration, that is possible if, either $h$ remains constant over time or $\Gamma$ remains constant over the entire span of the extra-dimension. Since the second possibility contradicts our assumption, we assume the first one to be true (i.e. $\dot{h}=0$, and we can consider $h=h(y)$ in effect) in the deductions that now follows. A quick look at equation (\ref{07c}) then reveals that, in such a case, the effective four-dimensional Newton's constant becomes fixed for our model, i.e. we are then dealing with a stabilized bulk.

Thus the field equations get simplified to the form
\begin{equation}\label{12}
\frac{3}{4\tau\Gamma}\left( \frac{2\dot{\tau}}{t\tau} + \frac{1}{t^{2}} + \left( \frac{\dot{\tau}}{\tau} \right)^{2} \right) + \frac{3}{4h}\left( \frac{\Gamma^{\prime}h^{\prime}}{\Gamma h} - \frac{2\Gamma^{\prime \prime}}{\Gamma} \right) = - \Lambda_{(5)} + 8 \pi G_{(5)} \bar{\rho},
\end{equation}

\begin{equation}\label{13}
\frac{1}{4\tau\Gamma} \left( \frac{4\ddot{\tau}}{\tau} + \frac{\dot{\tau}}{\tau} \left( \frac{1}{t} - \frac{3\dot{\tau}}{\tau} \right) - \frac{1}{t^{2}} \right) + \frac{3}{4h}\left( \frac{\Gamma^{\prime}h^{\prime}}{\Gamma h} - \frac{2\Gamma^{\prime \prime}}{\Gamma} \right) = - \Lambda_{(5)} - 8 \pi G_{(5)} \bar{P},
\end{equation}

and

\begin{equation}\label{14}
\frac{3}{4\tau\Gamma}\left( \frac{2\ddot{\tau}}{\tau} + \frac{3\dot{\tau}}{t\tau} - \left( \frac{\dot{\tau}}{\tau} \right)^{2} \right) - \frac{6}{4h}\left( \frac{\Gamma^{\prime}}{\Gamma} \right)^{2} = - \Lambda_{(5)} - 8 \pi G_{(5)} \bar{P}_{y}.
\end{equation}

In the calculations that now follow, we shall assume $8 \pi G_{(5)}=1$. From (\ref{12}) and (\ref{13}), we find that
\begin{equation}\label{14a}
\bar{\rho} + 3\bar{P} = \frac{3}{4\tau\Gamma}\left( - \frac{4\ddot{\tau}}{\tau} + \frac{2}{t^{2}} + \left( \frac{2\dot{\tau}}{\tau} \right)^{2} + \frac{\dot{\tau}}{t\tau} \right) - \frac{6}{4h}\left( \frac{\Gamma^{\prime}h^{\prime}}{\Gamma h} - \frac{2\Gamma^{\prime \prime}}{\Gamma} \right) - 2\Lambda_{(5)}
\end{equation}
and from (\ref{12})-(\ref{14}) we get
\begin{equation}\label{14b}
\bar{\rho} + 3\bar{P} + \bar{P}_{y} = \frac{3}{4\tau\Gamma}\left( \frac{2}{t^{2}} + 5\left(\frac{\dot{\tau}}{\tau}\right)^{2} - \frac{2\dot{\tau}}{t\tau} - \frac{6\ddot{\tau}}{\tau} \right) - \frac{6}{4h}\left( \frac{\Gamma^{\prime}h^{\prime}}{\Gamma h} - \frac{2\Gamma^{\prime \prime}}{\Gamma} - \left( \frac{\Gamma^{\prime}}{\Gamma} \right)^{2} \right) - 3\Lambda_{(5)}
\end{equation}

From the above it is evident that the validity of the strong energy condition in the bulk will be governed by the nature of warping, as well as by the effect of the extra dimension. However, the situation is too complicated at present for us to specify the nature of the constraints. The exact nature of these constraints can be determined only after some more analysis, which we consider in the next section.

\section{Solutions for the case where the Weyl tensor for the space-time vanishes}

Let us consider the five-dimensional bulk to be a space of constant curvature. We know that space-time metrics of constant curvature are locally characterized by the condition that the corresponding Weyl tensor vanishes \cite{HE}. The interesting feature of the metric given by equation (\ref{07}) is the symmetric nature of all the components of its Weyl tensor. The non-zero components are of the form

\begin{eqnarray*}
C^{A}_{BAB}= \frac{1}{24}\frac{\bar{g}_{BB}F(t)}{\tau\Gamma},\qquad for \qquad A=t;\qquad B=r,\theta,\phi \\
  \qquad \qquad\qquad or  \qquad  A=r, \theta, \phi; \qquad  B=y
\end{eqnarray*}

\begin{eqnarray*}
C^{A}_{BBA}= \frac{1}{8}\frac{\bar{g}_{BB}F(t)}{\tau\Gamma}, \qquad \qquad \qquad \qquad if \qquad \qquad \qquad A=y; B=t \\
  \qquad \qquad \qquad  = \frac{1}{24}\frac{\bar{g}_{BB}F(t)}{\tau\Gamma},\qquad if \qquad A=\theta; B=r  \qquad or \qquad A=\phi; B=\theta
\end{eqnarray*}
and similarly for all the other components, where
\begin{equation}\label{15}
F(t)=\frac{1}{t^{2}\tau^{2}}(-2t^{2}\tau\ddot{\tau} + t\tau\dot{\tau} + 3t^{2}\dot{\tau}^{2} + 2\tau^{2}).
\end{equation}

In general, we can write
\begin{equation}\label{16}
C_{ABAB}= constant \times \left(\frac{\bar{g}_{AA}\bar{g}_{BB}F(t)}{\tau\Gamma}\right),
\end{equation}
where, only the constant factor varies for the different components. Thus, for the case $F(t)=0$, the Weyl tensor for this space-time metric will vanish, thereby satisfying the condition for a constant curvature bulk. If we now impose the condition of isotropic pressure $\bar{P}=\bar{P}_{y}$, we get the result

\begin{equation}\label{17}
\frac{1}{\tau}\left( \left( \frac{1}{t} + \frac{\dot{\tau}}{\tau} \right)^{2} + \frac{\dot{\tau}}{t\tau} \right) - \frac{1}{h}\left( \Gamma^{\prime} \left( \frac{2\Gamma^{\prime}}{\Gamma} + \frac{h^{\prime}}{h} \right) - 2\Gamma^{\prime\prime} \right) = 0.
\end{equation}
Since the $\tau$ is a function of time and $h$ is a function of the fifth coordinate, the equation (\ref{17}) will be satisfied if the coefficient of $1/\tau$ and $1/h$ vanish separately.

The coefficient of $1/h$ will vanish when

\begin{equation}\label{18}
h=\left( \frac{1}{\Gamma} \frac{\partial\Gamma}{\partial y} \right)^{2},
\end{equation}

which implies that

\begin{equation}\label{19a}
\ln \Gamma(y)=\int h(y)^{1/2}dy.
\end{equation}

Thus $h$ is related to the $y$-dependent part of the warp factor. In view of equations (\ref{07a}) and (\ref{07b}), it follows that the effective four-dimensional Newton's constant $G_{N}$ depends on the effect of the extra dimension, i.e. the localization of gravity is determined by the effect of the extra dimension. If we now consider the solution
\begin{equation}\label{19b}
\Gamma(y)=e^{2l \ln(\cosh(cy))},
\end{equation}
then we obtain the extra-dimensional scale factor in the form
\begin{equation}\label{19c}
h^{1/2}=2lc\tanh(cy).
\end{equation}

From equation (\ref{17}), to have a vanishing coefficient for $1/\tau$ we need either $\tau$ to be negative or $\dot{\tau}$ to be so. Since the product $\tau\Gamma$ is positive and $\Gamma$ itself is also positive, $\tau$ must be positive in such a way that $\dot{\tau}$ is negative. In that case, the ratio $\dot{\tau}/\tau$ is negative, which may either be a constant or a function of time. For the case when this ratio is a negative constant, we obtain the solution
\begin{equation}\label{19d}
\tau=\frac{1}{C_{1}}e^{-\alpha t},
\end{equation}
$\alpha$ being a positive proportionality constant, $C_{1}$ is a constant of integration, which can be set equal to unity. At sufficiently late time, the universe must eventually accelerate. The acceleration is explained on the basis of the effect of a back reaction of the 5-dimensional warped geometry on the 3-brane \cite{maia4}.

At higher energies, we have particles escaping into the extra dimension, in which case $h=h(t,y)$, and we obtain a correlation between the extra-dimensional scale factor and both the time-dependent and the extra-dimension dependent parts of the warp factor. However, currently we confine to the low energy considerations.

Now, combining (\ref{12}) and (\ref{13}), we get
\begin{equation}\label{20}
\bar{\rho} + \bar{P} = \frac{3\dot{\tau}}{4t\tau^{2}\Gamma}.
\end{equation}

Since both $\tau$ and $\Gamma$ are positive and $\dot{\tau}$ is negative, the bulk matter will not obey the weak energy condition \cite{Wald}. This means that the matter in the bulk will be in the form of exotic matter. For the case of (\ref{19d}), the last equation will reduce to the form
\begin{equation}\label{21}
\bar{\rho} + \bar{P} = - \frac{3\alpha}{4t\tau\Gamma}.
\end{equation}

Further, we get
\begin{equation}\label{21a}
\bar{\rho} + 3\bar{P} + \bar{P}_{y} = \frac{9}{2h}\left( \frac{\Gamma^{\prime}}{\Gamma} \right)^{2} - 3\Lambda_{(5)} - \frac{21\alpha}{4t\tau\Gamma},
\end{equation}
fixing the condition for the strong energy condition to hold in the bulk.

\section{Evolution of the induced universe}

The induced metric is given by

\begin{equation}\label{23}
ds^2=\tau(t)\Gamma(y)\left(dt^2 -btdr^2 -btr^2d\theta^2 -btr^2sin^{2}\theta d\phi^2\right)=\tau(t)\Gamma(y)h_{\alpha \beta}(x)dx^{\alpha}dx^{\beta}.
\end{equation}
We find that the Weyl tensor for this metric is identically zero. The Ricciscalar turns out to be

\begin{equation}\label{23a}
R= - \frac{3}{2t\tau^{3}\Gamma}(2t\tau\ddot{\tau}+3\tau\dot{\tau}-t(\dot{\tau})^{2}).
\end{equation}

The expression for $R$ shows that the sign of $R$ is not positive definite. Since $\tau$ is positive and $\dot{\tau}$ is negative, the last two terms in the above expression will be negative. Therefore, the sign of $\ddot{\tau}$ and the magnitude of the first term will be crucial in determining the sign of $R$. The induced geometry will be a de Sitter one  when $R$ is positive and Anti de Sitter when $R$ is negative \cite{HE}.

The components of the extrinsic curvature for the hypersurface defined by (\ref{23}) are obtained as
\begin{center}
$K_{tt}=-\frac{1}{2}\tau\Gamma^{\prime}$,
\end{center}
\begin{center}
$K_{rr}=\frac{1}{2}\tau\Gamma^{\prime} bt$,
\end{center}
\begin{center}
$K_{\theta\theta}=\frac{1}{2}\tau\Gamma^{\prime} btr^{2}$
\end{center}
and
\begin{center}
$K_{\phi\phi}=\frac{1}{2}\tau\Gamma^{\prime} btr^{2}\sin^{2}\theta$
\end{center}
where, as mentioned earlier, prime denotes partial differentiation with respect to y. It is evident that the extrinsic curvature of this hypersurface is governed by the time-dependent warp factor. In compact form, we can write
\begin{equation}\label{24}
K_{\mu\nu}=-\frac{1}{2}\frac{\Gamma^{\prime}}{\Gamma}g_{\mu\nu}.
\end{equation}

Since $g_{\mu\nu}$ is diagonal, therefore, $K_{\mu\nu}$ also turns out to be diagonal. The time-dependent part of the warp factor is contained within $g_{\mu\nu}$.

The equations of motion for this braneworld embedded in a five-dimensional constant curvature bulk, can be derived directly from the Gauss and Codazzi equations for the conditions of integrability of the embedding geometry. The result is basically the Einstein field equations, modified by the presence of the extra term due to the extrinsic curvature as obtained by \cite{maia1},\cite{maia2} and is given by

\begin{equation}\label{25}
R_{\mu\nu}-\frac{1}{2}R g_{\mu\nu}+ \lambda g_{\mu\nu}=-8\pi G T_{\mu\nu} + Q_{\mu\nu},
\end{equation}
where, $\lambda$ denotes the effective cosmological constant in 4-dimension, including the vacuum energy.

The components of the 4-dimensional energy-momentum tensor are given by

\begin{center}
$T^{t}_{t}=\rho,\qquad\qquad and \qquad\qquad T^{i}_{j}=-p$.
\end{center}

For (\ref{25}) to be valid, we must have
\begin{equation}\label{26}
Q^{\mu\nu}_{;\nu}=0.
\end{equation}
Together with the conservation of $T_{\mu\nu}$, whatever is the influence of this term $Q_{\mu\nu}$ on the evolution of the brane, it turns out to be energetically uncoupled from the other components of the universe. The last term in equation (\ref{25}) is the contribution of the extrinsic curvature and is therefore purely geometrical, being given as

\begin{equation}\label{27}
Q_{\mu\nu}=g^{\rho\sigma}K_{\mu\rho} K_{\nu\sigma}-\xi K_{\mu\nu} - \frac{1}{2}(K^{2}- \xi^{2})g_{\mu\nu}\\=K^{\sigma}_{\mu}K_{\nu\sigma}-\xi K_{\mu\nu} - \frac{1}{2}(K^{2}- \xi^{2})g_{\mu\nu},
\end{equation}
where $\xi=g^{\mu\nu}K_{\mu\nu}$ denotes the mean curvature and $K^{2}=K^{\mu\nu}K_{\mu\nu}$ is the Gaussian curvature of the braneworld. Since
$Q_{\mu\nu}$ also depends on $g_{\mu\nu}$, it is also affected by the warp factor. Moreover, $Q_{\mu\nu}$ does not necessarily vanish, even when the bulk is flat. Therefore, it effectively modifies the usual dynamics of the gravitational field compared to that predicted from Einstein's theory. Since the effective cosmological constant depends on the scalar curvature of the bulk metric as well as $(\xi^2 - K^2)$ \cite{maia1} \cite{maia2}, hence the dynamics of the space-time should respond to the vacuum energy. For a constant curvature bulk, only the extrinsic curvature of the space-time corresponds to the vacuum energy. Thus, for the hypersurface under consideration, with the time-dependent nature of the extrinsic curvature, the vacuum energy is also a time-dependent entity. Therefore, as a consequence of considering a time-dependent warp factor for such warped product space-times, the vacuum energy of the corresponding braneworld turns out to be a time-dependent entity associated with the time-dependent process of bending of the braneworlds. Such braneworlds are characterized by the presence of a dark energy component.

In the remaining analysis, we assume that the sources on the brane is in the form of a perfect fluid satisfying a linear equation of state. Since we are in the low energy regime, we can neglect the contribution of the hot, relativistic matter called radiation and assume the matter content to be dominated by the cold, non-relativistic pressure-free matter, called dust (modelling for instance, the galactic fluid) and vacuum energy.

Equations (\ref{23}) and (\ref{24}) lead us to the result

\begin{center}
$\xi= -2 \left( \frac{\Gamma^{\prime}}{\Gamma} \right)$ \qquad and \qquad  $K^{2}=\left( \frac{\Gamma^{\prime}}{\Gamma} \right)^2$.
\end{center}
Thus both the Gaussian curvature and mean curvature of the hypersurface are constant for the $y=0$ hypersurface. Together with this, we get
\begin{equation}\label{28}
 Q_{\mu\nu}= \frac{3}{4}\left( \frac{\Gamma^{\prime}}{\Gamma} \right)^2 g_{\mu\nu}.
\end{equation}
The two non-trivial field equations are

\begin{equation}\label{29}
- \frac{3}{4\tau\Gamma} \left\{ \left( \frac{\dot{\tau}}{\tau} \right)^{2} + \frac{2\dot{\tau}}{t\tau} + \frac{1}{t^{2}} \right\} + \lambda = - 8\pi G_{N}\rho + \frac{3}{4}\left( \frac{\Gamma^{\prime}}{\Gamma} \right)^2
\end{equation}
and
\begin{equation}\label{30}
- \frac{1}{4\tau\Gamma} \left\{ \frac{4\ddot{\tau}}{\tau} + \frac{4\dot{\tau}}{t\tau} - \frac{1}{t^{2}} - \frac{3\dot{\tau}^{2}}{\tau^{2}} \right\} + \lambda = 8\pi G_{N}p + \frac{3}{4}\left( \frac{\Gamma^{\prime}}{\Gamma} \right)^2.
\end{equation}

\bigskip
The parameters $\rho$ and $p$ in the equations (\ref{29}) and (\ref{30}) are the total energy density and the total pressure of ordinary matter.
The effective energy density $\rho_{eff}$ of induced matter on the brane as obtained from (\ref{29}) is given by
\begin{equation}\label{31}
\rho_{eff}=\rho + \lambda - \frac{3}{4}\left( \frac{\Gamma^{\prime}}{\Gamma} \right)^2 = \rho + \lambda^{\prime},
\end{equation}
where, we have assumed that
\begin{center}
$8 \pi G_{N}=1$ \quad \quad and \quad \quad $\lambda^{\prime}= \lambda - \frac{3}{4}\left( \frac{\Gamma^{\prime}}{\Gamma} \right)^2 $.
\end{center}
Hence
\begin{equation}\label{31a}
\frac{3}{4\tau\Gamma} \left\{ \left( \frac{\dot{\tau}}{\tau} \right)^{2} + \frac{2\dot{\tau}}{t\tau} + \frac{1}{t^{2}} \right\} = \rho_{eff}.
\end{equation}

From (\ref{30}), we get the effective pressure of the induced matter as
\begin{equation}\label{32}
p_{eff} = p - \lambda + \frac{3}{4}\left( \frac{\Gamma^{\prime}}{\Gamma} \right)^2 = p - \lambda^{\prime},
\end{equation}
so that
\begin{equation}\label{32a}
\frac{1}{4\tau\Gamma} \left\{ \frac{1}{t^{2}} + \frac{3\dot{\tau}^{2}}{\tau^{2}} - \frac{4\dot{\tau}}{t\tau} - \frac{4\ddot{\tau}}{\tau} \right\}= p_{eff}.
\end{equation}

Combining (\ref{31}) and (\ref{32}), we obtain
\begin{equation}\label{33}
\rho_{eff} + p_{eff} = \rho + p
\end{equation}
and
\begin{equation}\label{34}
\rho_{eff} + 3p_{eff} = \rho + 3p - 2\lambda^{\prime}.
\end{equation}

For the case of dust, we get
\begin{equation}
\lambda= \frac{1}{4\tau\Gamma} \left( \frac{4\ddot{\tau}}{\tau} + \frac{4\dot{\tau}}{t\tau} - \frac{1}{t^{2}} - \frac{3\dot{\tau}^{2}}{\tau^{2}} \right) + \frac{3}{4}\left( \frac{\Gamma^{\prime}}{\Gamma} \right)^2 ,
\end{equation}
which shows that the effective cosmological constant for the induced matter is a variable quantity, monitored by the warp factor. Consequently, it evolves with time and also depends on the effect of the extra dimension.

Thus, the usual four-dimensional Friedmann equation relating the expansion rate $H$ of our universe with the energy density will get modified by the presence of the extra geometric term arising from the effect of warping. Although, in general, it is not necessary for the energy density on the brane to evolve as the square of the Hubble parameter, but with the induced metric in hand, we find that the evolution closely follows the standard predictions. In fact, the parameters may be suitably chosen so as to match well with the observations. From the expression of the scale factor, $A^{2}(\eta(t))= Cbt\tau(t)$ where $C=\Gamma(y=0)$, we obtain the Hubble parameter for the induced metric as
\begin{equation}\label{34a}
H = \frac{(\tau + t\dot{\tau})}{2t\tau}.
\end{equation}
Thus we have
\begin{equation}\label{34a1}
\rho = \frac{3H^{2}}{\tau\Gamma} + \frac{3}{4}\left( \frac{\Gamma^{\prime}}{\Gamma} \right)^{2} - \lambda.
\end{equation}
The last equation is very significant in the sense that when the warp factor is reduced to a constant, it resembles the usual results of standard cosmology. We also find that the expansion rate is not only determined by the ordinary matter on the brane, but also by the effect of the extra dimension. From (\ref{34a1}) we get
\begin{equation}\label{34a2}
\frac{\rho}{3H^2} + \frac{\lambda}{3H^2} - \frac{3}{12H^2}\left( \frac{\Gamma^{\prime}}{\Gamma} \right)^{2} = \frac{1}{\tau\Gamma},
\end{equation}
which is indeed the equation involving the density parameters of the various components in the universe. It is evident that the standard equation gets modified by the effect of the warp factor and the extra dimensional scale factor, contributing an additional geometric term to the total density parameter. When the induced universe will represent the observed universe, the sum of all such components of the density parameter should be equal to unity, thereby setting the constraint on the relation between $\tau$ and $\Gamma$.

\section{Cosmological consequences}
\label{sec:3}

\bigskip
From (\ref{33}), it is evident that if ordinary matter obeys the weak energy condition (WEC), i.e.

\begin{center}
$\rho + p \geq 0 $,
\end{center}
then the induced matter on the brane will also do the same, since in that case,
\begin{center}
$\rho_{eff} + p_{eff} \geq 0$.
\end{center}

Hence, both may be represented by a quintessence field. Further, to have $\rho_{eff} > 0$, we must have
\begin{equation}\label{34b}
\rho + \lambda_{eff} > 0,
\end{equation}
thereby setting a restriction on the value of $\lambda_{eff}$. For physically possible matter fields, $\rho > 0$, but unless $\lambda_{eff}$ satisfies appropriate conditions, we cannot have $\rho_{eff} > 0$. Moreover, even when $\rho_{eff} > 0$, we may have $p_{eff}<0$. In that case, assuming that the ordinary matter is represented by an equation of state of the form $p=w\rho$, we obtain from (\ref{32}),

\begin{equation}\label{34c}
w\rho - \lambda_{eff} < 0.
\end{equation}

From (\ref{34b}) and (\ref{34c}), we obtain the following restriction on the value of $\rho$:

\begin{equation}\label{34d}
- \lambda_{eff}< \rho < \frac{\lambda_{eff}}{w},
\end{equation}
for which the induced matter will experience negative pressure. Therefore, we conclude that, unless we have the ordinary matter as a phantom field, we cannot have the induced matter behaving as a phantom field, i.e. the behavior of the induced matter on the brane follows the behavior of the ordinary matter in the higher dimensional scenario in this case. The induced matter will be described by a negative pressure quintessence field provided $\rho$ satisfies the condition (\ref{34d}).

\bigskip
A careful examination of equation (\ref{34}) reveals the following facts:

\begin{itemize}
\item \textbf{Status of energy conditions}:Although ordinary matter in the higher-dimensional scenario may obey the strong energy condition (SEC), but that does not mean that the effective induced matter on the brane will also obey the SEC, unless appropriate conditions are satisfied by $\lambda_{eff}$. Hence for some value of $\lambda$ and $\Gamma$,  the induced matter on the brane may violate the SEC on the brane and behave as dark energy, although ordinary matter in the higher dimensional scenario will not behave as such. Such a thing will happen if $\rho + 3p \geq 0$, but $\rho + 3p < 2\lambda_{eff}$ i.e., $\rho + 3p < 2 \left[\lambda + \frac{3}{4}\left( \frac{\Gamma^{\prime}}{\Gamma} \right)^2 \right]$, so that $\rho_{eff} + 3p_{eff} < 0$. Since $\lambda$ is a variable quantity, monitored by the warp factor, the status of the energy conditions also depend on the effect of the warp factor. It must be noted here, that this strange behavior of the induced matter is basically a consequence of the embedding scheme.

\item \textbf{Deceleration parameter}:We know that for the Friedmann models, the two field equations together produce the result
\begin{center}
$\frac{\ddot{R}}{R}=\frac{\lambda}{3}-\frac{1}{6}(\rho+3p)$.
\end{center}
The deceleration parameter is given by
\begin{center}
$q= - \frac{\ddot{R}}{RH^{2}}$.
\end{center}
Therefore,
\begin{center}
$q=\frac{1}{6H^{2}}(\rho + 3p) - \frac{\lambda}{3H^{2}}$.
\end{center}
This means that, even if $\rho + 3p >0$, $q$ may not be positive because that will also depend on the contribution of the cosmological term, and accordingly there may be either acceleration or deceleration. However, if $\rho + 3p < 0$, then $q<0$ and the universe will exhibit acceleration. From the point of view of the induced matter on the brane, the induced matter will exhibit acceleration even when the matter in the higher-dimensional bulk may not exhibit acceleration.

The explicit form of the deceleration parameter for the induced metric is given by
\begin{equation}\label{36}
q = -1 - \frac{\dot{H}}{H^{2}} = -1 - \frac{2(t^{2}\tau^{3}\ddot{\tau} + t^{2}\dot{\tau}^{2} - \tau^{4})}{(\tau^{2} + t\tau\dot{\tau})^{2}}.
\end{equation}

From equation (\ref{34a}), we can say that initially, when $t$ was small, $H$ had a large value. That was the time before structure formation. But at later times, $H$ will decrease, because $\tau$ decreases with time. Finally, the evolution of the universe is dictated by the effect of the time-dependent warp factor, when the dynamics is dominated by the dark energy. Analyzing the nature of $q$ we can say that we shall have a universe which will be initially decelerated but will subsequently make a transition to an accelerated phase.
\end{itemize}
Therefore, the energy conditions as well as the behavior of the deceleration parameter depends strongly on the time-dependence of the warp factor and hence the dynamics of dark energy is also determined by it. The expression for the red-shift parameter $Z$ is obtained as

\begin{equation}\label{36a}
Z=1+z=\left( \frac{t_{0}\tau_{0}}{t\tau} \right)^{1/2},
\end{equation}
where $z$ is the standard redshift and the suffix '0' indicates the present values.

We note that the time-dependence of the scale factor is modified compared to the Friedman cosmology. Hence there may be alterations in the nucleosynthesis and structure formation in the present model, thereby modifying the age of the universe. However, one point in favour of this model is that the evolution of the energy density is of the order of the square of the Hubble parameter. Hence, we do not expect any significant departure from the standard results. Exact estimates can only be made in specific cases, one such being considered in the sequel.

\subsection{A Special Case}

\begin{figure}[h]
\epsfig{file=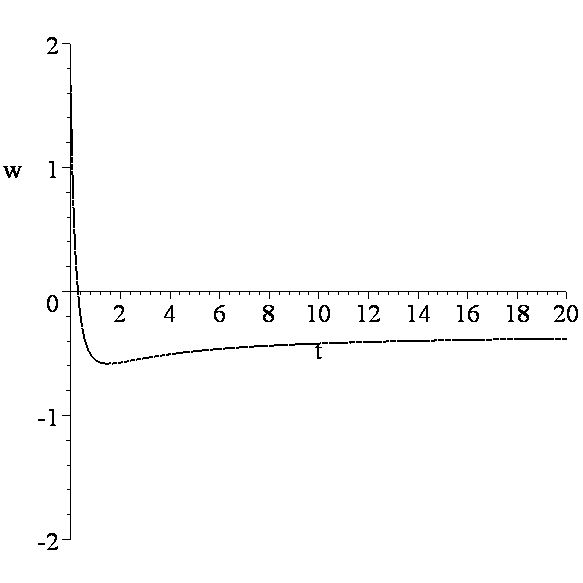,height=3.3in,width=5.5in}
\caption{Diagram showing evolution of $w$ with conformal time $t$ with the sample value of $a=1$. The diagram indicates that there is a transition from a decelerated phase to an accelerated phase.}
\label{state1}
\end{figure}

Let us consider the case for which $\tau(t)=e^{2at}$ with $\alpha=-2a$ in (\ref{18}) and $\Gamma(y)=e^{2l \ln(\cosh(cy))}$ \cite{GC}. In this case, the scale factor has a form suitable for representing the accelerated expansion at later times. The two field equations are now

\begin{equation}\label{39}
- \frac{3}{4e^{2(at + l\ln(\cosh(cy)))}}\left\{4a^{2} + \frac{4a}{t}+ \frac{1}{t^{2}}  \right\} + \lambda = - \frac{8\pi G\rho}{c^2} + \frac{3}{4}c^{2}\tanh^{2}(cy)
\end{equation}
and

\begin{equation}\label{40}
- \frac{1}{4e^{2(at + l\ln(\cosh(cy)))}}\left\{ 4a^{2} + \frac{8a}{t} - \frac{1}{t^{2}} \right\} + \lambda = \frac{8\pi G p}{c^{4}} + \frac{3}{4}c^{2}\tanh^{2}(cy),
\end{equation}
thereby yielding $\lambda_{eff}=\lambda - \frac{3}{4}c^{2}\tanh^{2}(cy)$. For the $y=0$ hypersurface, we have $\lambda_{eff}=\lambda$. The equation of state for the energy density is
\begin{equation}\label{40a}
w = - \frac{1}{3} - \frac{4at - 2}{ 4a^2t^2 + 4at + 1 }.
\end{equation}
From (\ref{40a}) we find that the universe will have a \emph{variable equation of state parameter} $w$. At later times we shall have $w<-1/3$, which is necessary for cosmic acceleration. Only for small values of $t$, for which $4at\ll2$, $w$ may have positive values. The variation of the equation of state parameter with the conformal time is shown in Fig.~\ref{state1}. The value of $w$ will fall from about $5/3$ in the very early universe around $t \sim 0$,  through the radiative phase (for which $w=1/3$), down to zero (dust-dominated phase) and then to negative values (vacuum energy dominated phase), crossing the mark of $-1/3$. It will decrease further and then rise slightly, attaining a nearly constant magnitude, suitable to sustain acceleration till very late in its evolutionary phase. At the present time when $t=t_{0}=1$, $w$ is surely $<-1/3$, indicating that in this model, the universe at present is in a state of accelerated expansion. However, $w$ does not reach the value $-1$, showing that the universe will not behave as a phantom field for this particular case even at late times.

The scale factor is a composite function $A^2=bte^{2(at+l)}$, where a power law type of variation is coupled with an exponential one. Thus at later times, the evolution will be dictated primarily by the exponential factor, modified by a half-power factor of conformal time, thereby determining the rate of formation of galaxies and the age of the universe.
\bigskip
The Hubble parameter and the deceleration parameter are now given by
\begin{equation}\label{41}
H= \frac{1+2at}{2t}
\end{equation}
and
\begin{equation}\label{42}
q = -1 + \frac{2}{(1+2at)^{2}}.
\end{equation}

Thus, during the initial stage of evolution of the universe, i.e. at small values of $t$, the hubble parameter has a large magnitude. However, at later times, when the dynamics is dominated by dark energy, $H$ will decrease and gradually settle down to a fixed value '$a$', which is the coefficient in the power of the time-dependent warp factor. The magnitude of '$a$' can be suitably chosen to fit with observational evidences. From (\ref{42}) we find that for acceleration, we must have

\begin{center}
$at>0.212$.
\end{center}

\begin{figure}[h]
\epsfig{file=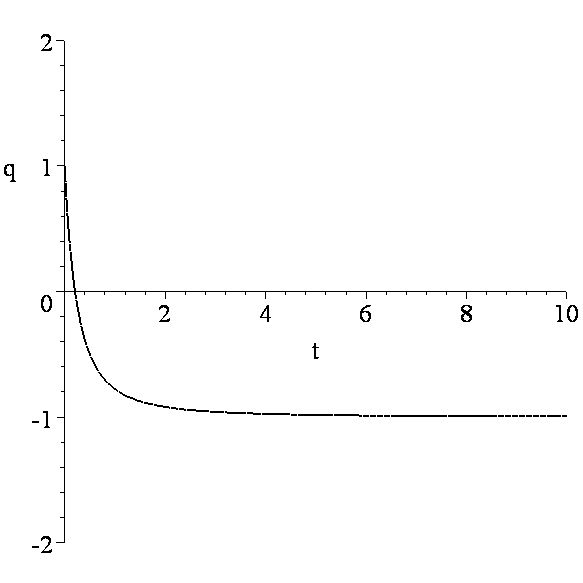,height=3.3in,width=5.5in}
\caption{Diagram showing evolution of $q$ with conformal time $t$ with the sample value of $a=1$. The transition from decelerated phase to the
accelerated phase is clearly evident.}
\label{state2}
\end{figure}
\begin{figure}[h]
\epsfig{file=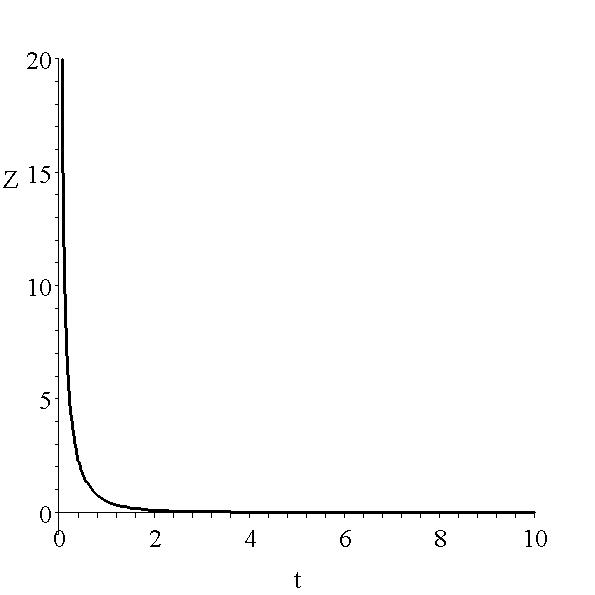,height=3.3in,width=5.5in}
\caption{Diagram showing evolution of $Z$ with time with the sample value of $a=1$ and $t_{0}=1$.}
\label{state3}
\end{figure}

Fig.~\ref{state2} shows a plot of $q$, which indicates this result. The redshift parameter $Z$ is now
\begin{equation}\label{42a}
Z=\left( \frac{t_{0}}{t}\right)^{1/2}e^{a(t_{0}-t)}.
\end{equation}
The variation of redshift with the conformal time $t$ is indicated in Fig.~\ref{state3} with $t_{0}=1$ representing the present time.

\bigskip
The conformal age of the universe is given by
\begin{equation}\label{42b}
t=\int\frac{dt}{(bt)^{1/2}e^{(at+l)}}.
\end{equation}

\subsubsection{A model of dark energy measured by the age of the universe}

Writing down the metric in (\ref{18}) in terms of the cosmic time $\eta(t)$, with  for the $y=0$ hypersurface we have
\begin{equation}\label{43}
ds^{2}=d\eta^2 -a^2\eta^2\frac{b}{a}\ln(a\eta)(dr^2 + r^2d\theta^2 + r^2sin^{2}\theta d\phi^2).
\end{equation}
Defining $a\eta=\eta^{\prime}$, we obtain the above metric in the form
\begin{equation}\label{44}
ds^{2}=\frac{1}{a^2}(d\eta^{\prime})^2 - (\eta^{\prime})^2\frac{b}{a}\ln\eta^{\prime}(dr^2 + r^2d\theta^2 + r^2sin^{2}\theta d\phi^2).
\end{equation}
The expansion parameter is now
\begin{equation}\label{45}
A^2(\eta^{\prime})=(\eta^{\prime})^2\frac{b}{a}\ln\eta^{\prime}
\end{equation}
and the Hubble parameter is
\begin{equation}\label{46}
H = \frac{2\ln\eta^{\prime} + 1}{\eta^{\prime}\ln\eta^{\prime}}.
\end{equation}

The components of the Einstein tensor are now
\begin{equation}\label{47}
G^{\eta^{\prime}}_{\eta^{\prime}}=\frac{3a^2(4\ln(\eta^{\prime})^2 + 4\ln\eta^{\prime} + 1)}{4(\eta^{\prime})^2\ln(\eta^{\prime})^2}
\end{equation}
and
\begin{equation}\label{48}
G^{i}_{j}=\frac{a^2(4\ln(\eta^{\prime})^2 + 8\ln\eta^{\prime} - 1)}{4(\eta^{\prime})^2\ln(\eta^{\prime})^2}.
\end{equation}
In view of (\ref{46}), we can write (\ref{47}) as
\begin{equation}\label{49}
G^{\eta^{\prime}}_{\eta^{\prime}}=\frac{3a^2H^2}{4}.
\end{equation}
In other words,
\begin{equation}\label{50}
H^2=\frac{4G^{\eta^{\prime}}_{\eta^{\prime}}}{3a^2},
\end{equation}
and the corresponding field equation is
\begin{equation}\label{51}
\frac{3a^2H^2}{4}=\rho+\lambda=\rho+\rho_{\lambda}.
\end{equation}

In this context, we note that recently Cai \cite{Cai} proposed a model of dark energy, in which the energy density of quantum fluctuations in the universe was interpreted as the dark energy present in our universe. The energy density for the quantum field was given as

\begin{equation}\label{52}
\rho_{q}=\rho_{\lambda}=\frac{3n^2m_{p}^{2}}{T^2},
\end{equation}
where $T$ is the age of the universe, $m_{p}$ is the reduced Planck mass and $n$ is a numerical factor of order one, so as to match with observational data. Particularly, the energy density could drive the universe to accelerated expansion if $n>1$. The energy density (\ref{52}) with the current age of the universe, $T\sim1/H_{0}$ (where $H_{0}$ is the current Hubble parameter of the universe), gave a fairly accurate estimate of the observed dark energy density. In that case, the Friedman equation for a flat FRW universe with dust matter coupled with the quantum field was

\begin{equation}\label{53}
3m_{p}^{2}H^2=\rho+\rho_{\lambda}.
\end{equation}
We see that equations (\ref{51}) and (\ref{53}) are very much similar. Thus we can conclude that the present model represents a model of dark energy with the age of the universe as the measure of length, driving the universe to a state of accelerated expansion at later times.

The cosmic age of the universe can be calculated from
\begin{equation}\label{54}
T=\int\frac{dA(\eta^{\prime})}{H(\eta^{\prime})A(\eta^{\prime})}.
\end{equation}

\section{Summary and conclusions}
\label{sec:5}

This paper deals with a five-dimensional warp product space-time having time-dependent warp factor and a non-compact fifth dimension. We know that the warp factor reflects the confining role of the bulk cosmological constant to localize gravity at the brane through the curvature of the bulk. This process of localization may include some time-dependence. Hence we have considered a warp factor which depends both on time as well as on the extra coordinate. The extra dimensional scale factor is also a function of time and of the extra coordinate. Consequently, the field equations became very complicated. To simplify our analysis, the solutions were sought in the low energy regime (within the TeV scale), for which the geometry was found to represent a stabilized bulk. Further we assumed the bulk to be a space of constant curvature and the warp factor in the product form: a function of time and a function of the extra-dimensional coordinate $y$. The braneworld was defined by a flat FRW-type metric in the ordinary spatial dimension. The solutions for such a bulk, with isotropic pressure, was obtained. We have found that the y-dependent part of the warp factor depends on the metric coefficient of the extra dimension. The matter in the bulk is in the exotic form, which violates the strong energy condition.

For the induced metric, the extrinsic curvature was found to be governed by the time-dependent warp factor. We have obtained the brane field equations in the usual Einstein form with one extra term in the matter component. This extra matter term however, depends on the geometric quantities, namely the extrinsic curvature, the mean curvature and the metric coefficients. Thus the effective cosmological constant on the brane is a variable quantity, monitored by the warp factor, leading to a geometric interpretation of dark energy. Although the standard equation for the density parameter is modified by the effect of the warp factor and the extra dimensional scale factor, in spite of that, the evolution closely follows the standard predictions. The effective matter on the brane is related to the ordinary matter as follows: If the ordinary matter satisfies (violates) the weak energy condition, then the effective matter also does so, but for strong energy condition, the situation may not be identical. Depending on the warp factor, the effective matter may violate the strong energy condition but the ordinary matter obeys it. Finally, we have presented a cosmological solution as an example. Here the universe possesses a variable equation of state parameter, driving it through the radiative phase and the matter dominated phase, finally into the vacuum energy dominated phase. The evolution in terms of cosmic time represents a model of dark energy with the age of the universe as the measure of lengthdriving the universe to a state of accelerated expansion at later times.

In this paper, we have worked with a simple type of time-dependent warp factor. The consequences for a more generalized type of such a warp factor are being determined, which will be reported in a future work.

\section*{Acknowledgments}
A part of this work was done in IUCAA, India under the associateship programme. SG gratefully acknowledges the warm hospitality and the facilities of work at IUCAA. SC is thankful to CSIR, Govt. of India for funding a project (03(1131)/08/EMR-II).


\begin{thebibliography}{}
\bibitem{kk}T. Kaluza, \textit{Zum Unit\"{a}tsproblem der Physik}, Sitz. Preuss. Akad. Wiss. Phys. Math. \textbf{K1} 966 (1921). English translation in \textit{Unified field theories of more than four dimensions}, Proc. Int. School of Cosmology and Gravitation (Erice) edited by V. De Sabbata and E. Schmutzer (World Scientific, Singapore, 1983)
\bibitem{rbvspv}V. A. Rubakov and M. E. Shaposhnikov, Phys. Lett. \textbf{125B}, 136 (1983)
\bibitem{aadd}N. Arkani-Hamed, S. Dimopoulos and G. Dvali, Phys. Lett B \textbf{429}, 263 (1998); I. Antoniadis, N. Arkani-Hamed, S. Dimopoulos, G. Dvali, Phys. Lett B \textbf{436}, 257 (1998)
\bibitem{add}N. Arkani-Hamed, S. Dimopoulos and G. Dvali, Phys. Rev. D \textbf{59}, 086004 (1999)
\bibitem{khoury} J. Khoury, B. A. Ovrut, P. J. Steinhardt, and N. Turok, Phys. Rev. D64, 123522 (2001)
\bibitem{rs1}L. Randall and R. Sundrum, Phys. Rev. Lett. \textbf{83}, 3370 (1999)
\bibitem{rs2} Randall L and Sundrum R 1999 { Phys. Rev. Lett.} { 83} 4690
\bibitem{lrr} Maartens R 2004 { Liv. Rev. Rel.} 2004-7 (http://www.livingreviews.org/lrr-2004-7)
\bibitem{maia1} Maia M D, Monte E M and Maia J M F 2004 { Phys. Lett. B} { 585} 11
\bibitem{maia2} Maia M D, Monte E M, Maia J M F and Alcaniz J S 2005 { Class. Quant. Grav.} { 22} 1623
\bibitem{Nihei} Nihei T 1999 { Phys. Lett. B} { 465} 81
\bibitem{bdl} Binetruy P, Deffayet C and Langlois D 2000 {\it Nucl. Phys. B} {\bf 565} 269\\ Binetruy P, Deffayet C, Ellwanger U and Langlois D 2000 {\it Phys. Lett. B} {\bf 477} 285
\bibitem{GC}Such a warping function has been already considered by the authors. We know that the bulk cosmological constant acts to "squeeze" the gravitational field closer to the brane, with the exponential warp factor reflecting the confining role of the bulk cosmological constant \cite{lrr}. The process of localization may include some time-dependence. Mathematically, the time dependence of the warp factor does not affect the smooth nature of the scalar function $f$. Physically, it takes into account the possibility that the confining role of the bulk cosmological constant and the curvature of the bulk may have some dependence on time, in addition to their dependence on the extra dimensional coordinate and hence the choice. See Guha S and Chakraborty S 2010 {Gen. Relativ. Grav.}; arXiv:0812.5072
\bibitem{maia4} Maia M D and Silva G S 1994 { Phys. Rev. D} { 50} 7233
\bibitem{dahia3} Dahia F, Romero C, Silva L F P and Tavakol R 2007 {\it J. Math. Phys.} {\bf 48} 072501
\bibitem{dahia4} Dahia F, Romero C, Silva L F P and Tavakol R 2008 {\it Gen. Rel. Grav.} {\bf 40} 1341
\bibitem{garriga} Garriga J and Tanaka T 2000 { Phys. Rev. Lett.} {84} 2778
\bibitem{giddings} Giddings S B, Katz E and Randall L 2000 { JHEP} {03} 023
\bibitem{ant1} Antoniadis I, Phys. Lett. B 246 (1990) 377
\bibitem{ant2} Antoniadis I, C. Munoz, M. Quirós, Nucl. Phys. B 397 (1993) 515; I. Antoniadis, K. Benakli, M. Quirós, Phys. Lett. B 331 (1994) 313
\bibitem{Lykken} Lykken J D, 1996 { Phys. Rev. D} { 54}  3693
\bibitem{Chung} Chung D J H and Freese K 1999 { Phys. Rev. D} { 61} 023511
\bibitem{clinevinet} Cline J and Vinet J 2003 { Phys. Rev. D} { 68} 025015
\bibitem{ant3} I. Antoniadis, S. Dimopoulos, A. Pomarol, M. Quirós, Nucl. Phys. B 544 (1999) 503
\bibitem{ponton}E. Ponton, E. Poppitz, JHEP 06 (2001) 019
\bibitem{HE} Hawking S W and Ellis G F R 1973 {\it The large scale structure of space-time} (Cambridge University Press)
\bibitem{Wald} Wald R M 2006 {\it General Relativity} (Overseas Press (India) Pvt. Ltd.)
\bibitem{Cai} Cai R G 2007 { Phys. Lett. B} { 657} 228
\end{thebibliography}
\end{document}